\documentstyle[epsf,12pt]{article}
\begin{document}
\begin{titlepage}
\begin{center}
\vspace*{4cm}

\begin{title}
\bold {\Huge 
On the Bose-Einstein Effect and the W Mass
}

\end{title}
\vspace{2cm}

\begin{author}
\Large {K. FIA{\L}KOWSKI\footnote{e-mail address:
uffialko@thrisc.if.uj.edu.pl} and R. WIT }
\end{author}\\
\vspace{1cm}
{\sl Institute of Physics, Jagellonian University \\
30-059 Krak{\'o}w, ul.Reymonta 4, Poland}

\vspace{3cm}
\begin{abstract}
   We present an implementation of the Bose - Einstein effect in a Monte Carlo
generator for $W^-W^+ $  production in the $e^-e^+$ annihilation by means
of the weight method. We check that the shift of the W mass in four jet events 
due to this effect is similarly
small as for the other prescription used recently by Jadach and Zalewski.
Possible generalization of this result is shortly discussed.
\end{abstract}
\end{center}
\end{titlepage}

\section{Introduction}
\par
The problem of an exact determination of the W mass is crucial for any test of
the standard model.  Thus all the effects resulting in the W mass shift are of great
importance. In particular, the possible mass shifts from the Bose - Einstein effect
 in the four jet events from the $e^-e^+$ annihilation into  $W^-W^+ $ pairs
 have been recently estimated with widely varying results [1, 2, 3].
\par
In this note we use the implementation of the Bose - Einstein effect based on
the Bia\l{}as - Krzywicki prescription for weights to be attached to the Monte
Carlo generated events [4]. We avoid the prohibitive increase of computational
time with multiplicity by the approximation used already to describe the effect
of hadronic collisions [5]. We follow closely the recent investigation by
Jadach and Zalewski [3] to check if  a different prescription for the weights
influences the physical conclusions.
\par
We describe our method shortly in the next section stressing the similarities
and differences with other approaches. The results are presented in the third
section. The last section contains discussion and conclusions.

\section{Implementation of the  weight method for the Bose - Einstein effect}
\par
The original discussion of the Bose - Einstein effect in multiparticle
production assumed the knowledge of amplitudes which have to be symmetrized 
[6]. This is not the case for Monte Carlo - based models, where only probabilities
are calculated. On the other hand, the multitude of available data can be
analyzed only within the models using Monte Carlo generators. This sort of
analysis seems to be necessary to discriminate among various pictures of the
space time developement of multiple production.  
Therefore various methods have been devised to implement the effect
into Monte Carlo generators. For the heavy ion collisions one used the semi -
classical description of the process to provide the space time distribution of
sources producing plane waves to be symmetrized [7]. For the Lund model of 
two-jet processes there is a natural measure for probabilities after symmetrization
[8]. However, the most widely used method imitates the effect just by suitable
shifts of final state momenta to get the experimental two - particle
distributions [9]. Apart from other problems this method includes the momentum
rescaling which results in serious mass shifts for W bosons [1,3]. Thus there is a
need for a reliable and general method to implement the Bose - Einstein effect
into any Monte Carlo generator. 
\par
Such a method seems to be suggested by the analysis by means of Wigner functions
pioneered by Pratt [10] and recently recalled by Bia\l{}as and Krzywicki [4].  We
have described  this method in Ref. [5]. For reader's convenience we repeat here
its main ingredients. 
\par
 With few simplifying assumptions one arrives at formulae where the multiparticle
density distribution is expressed by a product of original (non - symmetrized)
distribution and the weight factor, representing the effect of symmetrization.
In this way one gets a simple prescription for the Monte Carlo generators: one
should generate events according to the original generator and then attach to
each event its weight  calculated from a simple formula

\begin{equation}
\label{s1}
W(n) = \sum_{\{P(k)\}}\prod_{i=1}^{n}w_{iP(i)}.
\end{equation}
 Here $n$ is the number of identical particles, $w_{iP(i)}$ is a two particle weight factor 
calculated for the pair of momenta (of the $i-th$ particle and the particle which occupies 
the $i-th$ place in the permutation ${P(k)}$). The sum extends over all the
permutations of $n$ elements.
 Since all factors are positive and $w_{ii} =1$,
the resulting weight is not smaller than one (a contribution from identity
permutation). One may rescale the weights to keep, e.g., the average number of
particles fixed; we return to this point later. 
\par
Since most of the particles detected in experiments are pions, the final weight
should be actually given by a product of weights calculated separately for
positive, negative and neutral pions.  In fact, the  BE interference for neutral
particles is not observable (apart from the possible effects for direct photons
[11]): neutral pions decay before
detection, and for the resulting photons the effective source size is so big
that the BE effects must be negligible for momentum differences above a few eV.
However, the procedure should not change the observable correlations between the
numbers of charged and neutral pions. Therefore weights for all signs of pions
must be taken into account.
\par
Thus in principle the only arbitrary factor is the function of the difference of
two momenta $w_{ij}(p_i -  p_j)$.  As in Ref. [5]  we use here simply 
the Gaussian function of four-momentum difference squared 
\begin{equation}
\label{s2}
w_{ij} = e^{(p_i - p_j)^2/2\sigma ^2}
\end{equation}
which is motivated by a commonly used experimental parametrization of BE
effects.   
\par
Of course,  different components of momentum difference squared may be
multiplied by different coefficients, and the shape may be modified. In this
note we do not discuss these possibilities. Therefore the only parameter is a
Gaussian half-width of the distribution $\sigma$.
\par 
Unfortunately, for more than ten pions of a given sign the calculations become
prohibitively long. Symmetrizing separately in hemispheres [12] one shifts only
the problem to higher energies. There exists a scheme for calculating the sum
(1) in a reasonable time [13], but the method is still under investigation.  
In our calculations we have separated the sum of all the  n! permutations 
into terms where only the permutations which change places of exactly $K$ particles
are taken into account:
\begin{equation}
\label{s3}
w = \sum_Kw^{(K)}.
\end{equation}

The higher terms in this expansion correspond to configurations where many
particles have approximately the same momenta, which is very unlikely. 
These terms for $K<6$ are

$$
\begin{array}{lll}
w^{(0)} = 1; \   w^{(1)} = 0; \  
w^{(2)} = \sum_{i=1}^{n-1}\sum_{j>i}(w_{ij})^2; \ 
w^{(3)}  = 2\sum_{i=1}^{n-2}\sum_{j>i}\sum_{k>j}w_{ij}w_{jk}w_{ki}; & &
\\
 & & \\
w^{(4)} = 
\sum_{i=1}^{n-3}\sum_{j>i}\sum_{k>j}\sum_{l>k}[2w_{ij}w_{ik}w_{jl}w_{kl}
+2w_{ij}w_{il}w_{jk}w_{kl} +2w_{ik}w_{il}w_{jk}w_{jl} + (w_{il}w_{jk})^2 + & &
\\
 & &  \\ 
(w_{ij}w_{kl})^2 + (w_{ik}w_{jl})^2]; & & 
\\
 & & \\
w^{(5)} =
2\sum_{i=1}^{n-4}\sum_{j>i}\sum_{k>j}\sum_{l>k}\sum_{m>l}[(w_{ij})^2w_{lk}w_{ml}w_{km}
+(w_{ik})^2w_{jl}w_{ml}w_{jm} + &&  
\\
\end{array}
$$
\begin{equation}
\label{s4}
\begin{array}{lll}
(w_{il})^2w_{jk}w_{jm}w_{km} + (w_{im})^2w_{jk}w_{kl}w_{jl} + (w_{jk})^2w_{il}w_{lm}w_{im}
+(w_{jl})^2w_{ik}w_{km}w_{im} +  & &
\\
 & &\\
(w_{jm})^2w_{ik}w_{kl}w_{il} + (w_{kl})^2w_{ij}w_{jm}w_{im} +(w_{lm})^2w_{ij}w_{jk}w_{ik} +
 (w_{km})^2w_{ij}w_{jl}w_{il} +  && \\ 
&& \\
w_{ij}w_{jk}w_{kl}w_{lm}w_{im} +  w_{ik}w_{jl}w_{km}w_{jm}w_{il} +w_{il}w_{ij}w_{kl}w_{jm}w_{km} + 
w_{ij}w_{ik}w_{jl}w_{lm}w_{km} + && \\
&& \\w_{ik}w_{im}w_{jk}w_{jl}w_{lm} +  
w_{il}w_{jl}w_{jk}w_{km}w_{im} + w_{ij}w_{ik}w_{kl}w_{lm}w_{jm} + 
w_{ij}w_{il}w_{lm}w_{jk}w_{km} + && \\
&& \\
w_{ij}w_{im}w_{jl}w_{km}w_{kl} + 
w_{ik}w_{il}w_{jk}w_{lm}w_{jm} +w_{ik}w_{im}w_{jm}w_{jl}w_{kl} +
 w_{il}w_{im}w_{kl}w_{jk}w_{jm}].
\end{array}
\end{equation}
\par

The shape of the weight factor (2) should be chosen to fit the "BE ratio", defined for the pair 
of identical pions as a function of $Q = \sqrt {-(p_1 -p_2)^2}$
\begin{equation}
c_2(Q) = \frac {\int d^3p_1d^3p_2 \rho _2(p_1,p_2)\delta [Q-
\sqrt{-(p_1-p_2)^2}]} {\int d^3p_1d^3p_2 
\rho _1(p_1)\rho _1 (p_2)\delta [Q-
\sqrt{-(p_1-p_2)^2}]} \frac {<n>^2}{<n(n-1)>}.
\end{equation}
\par
Without weights it is rather flat and close to one for typical Monte Carlo
generators, if we normalize separately the numerator and the denominator 
of Eq. (5) to the same number of entries
(which is  achieved by the second factor  in (5)).  Including weights
produces a maximum at smallest $Q^2$  with the height about 2 (i.e. one unit
above the value at large $Q^2$) and a width $\sigma '$ close to $\sigma $ of
formula (2). Thus we reproduce satisfactorily the shape assumed for the 
two-particle weight factor.
\par
We have checked in Ref. [5] for PYTHIA/JETSET generated $p\overline{p}$ events
at 630 GeV that cutting the series (3)
at $K=3$ and at $K=4$  we get quite similar shapes of the $Q^2$
spectra, although the normalization is significantly different. Including
the term with $K=5$ we change even less all the  distributions. Thus we feel that 
cutting the series (3) at $K=5$ we 
get a reliable estimate of the results for $Q^2$  spectra from the weight method
(up to the possible change of normalization).
\par
This may seem surprising if we remember that our approximation does not take
into account, e.g., the contribution from such a simple configuration as three
pairs of very close (pairwise) momenta. Indeed, in this case there is a contribution from a
permutation of 6 elements. However, the full contribution of such a
configuration to the sum (1) is equal 1+3+3+1 = 8 (from permutations moving 0,
2, 4 and 6 elements, respectively) and our approximation counts all but the last
term in this sum.  We have checked that for all reasonably probable configurations
our approximation seems similarly satisfactory. 
\par
Since the JETSET/PYTHIA parameters were fitted to reproduce inclusive
experimental data without weights, the change, e.g., of the average multiplicity induced by
weights should be compensated by the proper 	refitting procedures. Instead we have
applied (as in Ref. [3]) a simple method of multiplying weights by an extra $cV^n$
factor, where n is the number of all "direct"  pions, and $c$ and $V$ are constants
fixed by the requirements to restore the original number of events and the
original average multiplicity. This is done by assuming  
that the original multiplicity distribution of "direct" pions may be well
approximated by the negative binomial formula, i.e. that the NBD parameters
$\overline n$ and $1/k$ are given by the experimental values of $<n>$ and $<n(n-1>/<n>^2
-1$.
If with the weights we get a new average multiplicity $<n'>$, the original
value may then be restored by rescaling the weights with 
\begin{equation}
V = \frac{<n>(<n'>+k)}{<n'>(<n>+k)}
\end{equation}
 and 
\begin{equation}
c =\frac {[1+(1- V)<n'>/k]^k}{<w>},
\end{equation}
where $<w>$ is the average value of weights before
rescaling. We have checked that this procedure restores indeed the original
 average multiplicity with accuracy of few percent. If this accuracy is not
satisfactory, the quantities $c$ and $V$ can be estimated by direct minimization
of differences between the multiplicity distributions without weights and with the
rescaled weights, respectively. On the other hand, the BE
ratios are little affected by rescaling (only the normalization, which is anyway
mainly a matter of convention, changes by a few percent). As will be shown
later this is also true for $e^-e^+$ collisions.
The BE ratio still reflects mainly the assumed shape of the
two-particle weight (plus 1): for larger $\sigma$ it is wider and starts to increase 
above 2 for smallest  $Q^2$.
\par
The procedure seems to produce  too high a value of the BE ratio for
smallest  $Q^2$. As already noted, it is about twice the value for
large $Q^2$, whereas in most of the data it is only by some 50\% higher. 
To explain why the BE ratio does not increase up to the value of 2, one may invoke 
some coherent component [6], but a more obvious effect (which also lowers 
the BE ratio) is the existence of longer living resonances. Pions coming
from their decay are effectively "born" more than 10 fm from the collision
point. Thus the Gaussian width parameter in a two-particle weight for these
pions should be smaller by an order of magnitude, which allows practically to
neglect their contribution to the BE effect in the experimentally accessible  $Q^2$
range. Therefore the Bia{\l}as - Krzywicki weights are calculated taking
into account only the permutations  of momenta of pions produced directly, or
resulting from the decay of the widest resonances.

\section{Results and comparison with data}
\par
We have generated 150 000 events of $e^+e^-$ annihilation into $W^+W^-$  
at 172 GeV CM energy by the
default version of the PYTHIA/JETSET generator [9]. For
each event the weight factor was calculated by taking the 4-momenta of "direct"
pions of each sign, calculating for them a matrix of two - particle weights
$w_{ij}$ according to (2) with $\sigma  = 0.14$ GeV (the same value as used to
describe the $p\overline p$ collisions), and then the weight $w$ as a
series (3) cut at $K = 5$ and rescaled as described in the previous 
section\footnote {More precisely, we have used the values of $c$ and $V$ fitted to 
restore the original multiplicity distribution for the hadronic decay of a single $W$.}. 
As already noted, the event weight is a product of
 weight factors for all three kinds of pions.
In Fig.1 we present the  ratio of "BE ratios" (5) for pairs of positive
pions as a function of $Q^2$ for the events from our prescription with series (3) 
cut at $K=5$ and from the standard  PYTHIA/JETSET generator (without weights). 
Results are shown for rescaled and unrescaled weights.
\vspace{0.5cm}

\epsfxsize=10cm

~~~~~~~~~~~~~~~\epsfbox{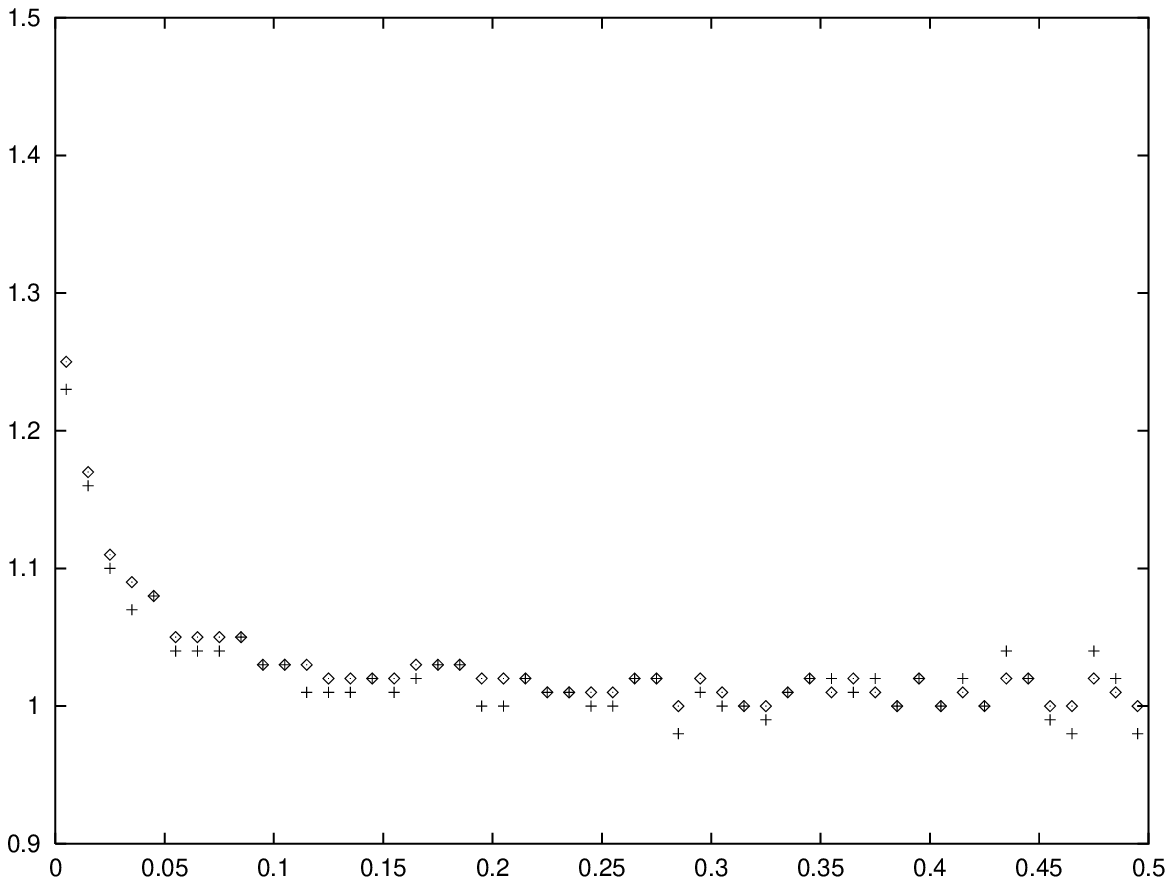}

\vspace{0.5cm}
\par
 {\bf Fig.1.} {\sl The  ratio of   "BE ratios" (5) for positive pions with - and
without weights  as a
function of  $Q^2 \  [GeV^2]$. Diamonds and crosses correspond to rescaled and unrescaled 
weights, respectively.}  

\vspace{0.5cm}

We see that without any fitting  we reproduce the main features of inclusive
hadroproduction data at similar energy [14] (there are no data yet for $W^+W^-$
events) and that the two curves are hardly distinguishable. 
\par
Next we analyze the events using the LUCLUS procedure
[15] to select 4 jet events and to
reconstruct W as the 2-jet system. To reduce the combinatorial background we
select as a partner for the first jet this jet which maximizes the two-jet
invariant mass. Then, as in Ref. [3], we plot the average of this mass and the
mass of the system of two remaining jets. The resulting W mass distribution with- and
without weights is shown in Fig.2.
We see that the introduction of weights hardly affects the distribution. The
average mass for two curves of Fig. 2 differs by less than 20 MeV. If we fit
these curves by the Breit - Wigner formula (plus background), the fitted $W$
mass would differ even less.
\vspace{0.5cm}
\epsfxsize=10cm

~~~~~~~~~~~~~~~\epsfbox{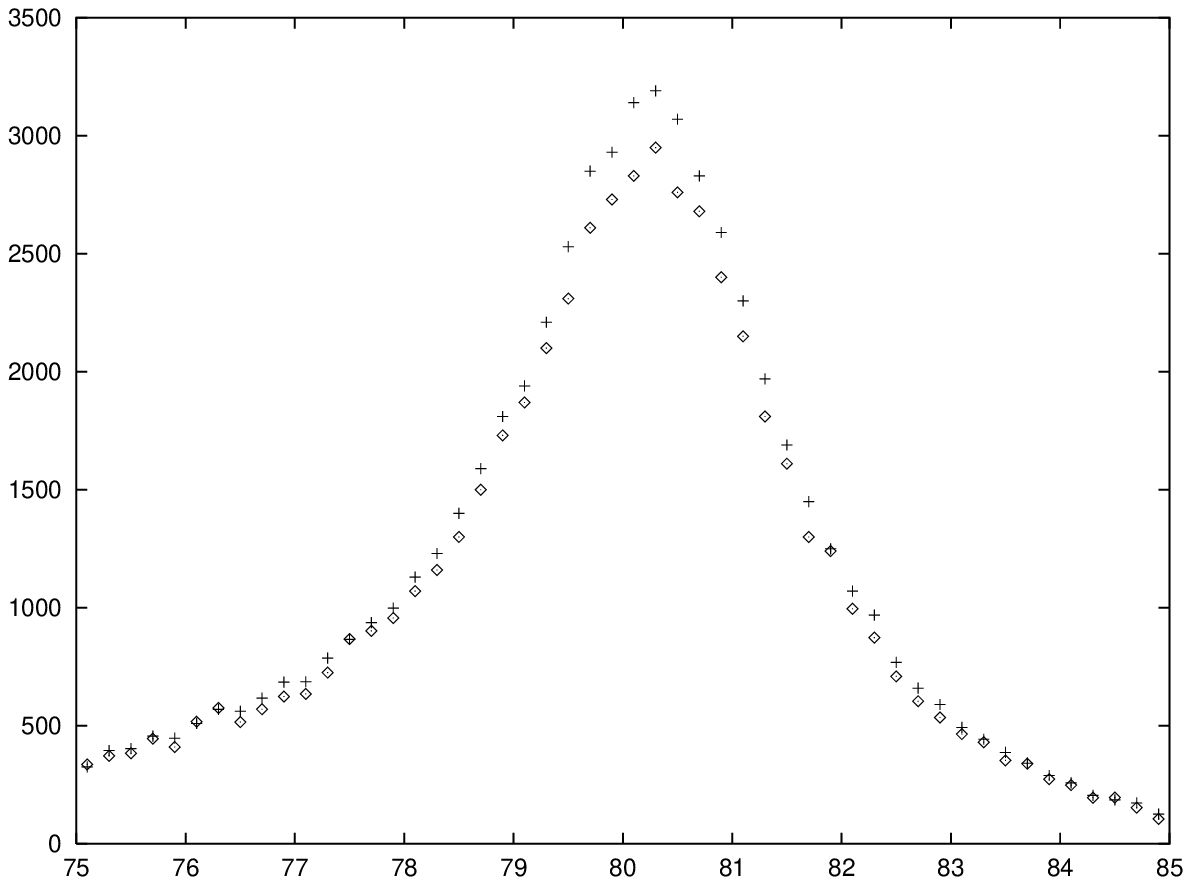}

\vspace{0.5cm}
\par
 {\bf Fig.2.} {\sl The two-jet invariant mass distribution (in $GeV/c^2$) as described in the text.
Diamonds and crosses correspond to distributions with - and without weights.}   \\

\par
It is important to underline that our results are very similar to those of Ref.
[3] both for the BE ratio and for the $m_W$ distribution, although our
distribution of weights has a much more extended tail, even after rescaling.
As shown in Fig. 3,  this tail decreases quite slowly and for 175 events the
values of weights are outside the plot (the maximal value is about hundred).

\vspace{0.5cm}
\epsfxsize=10cm

~~~~~~~~~~~~~~~\epsfbox{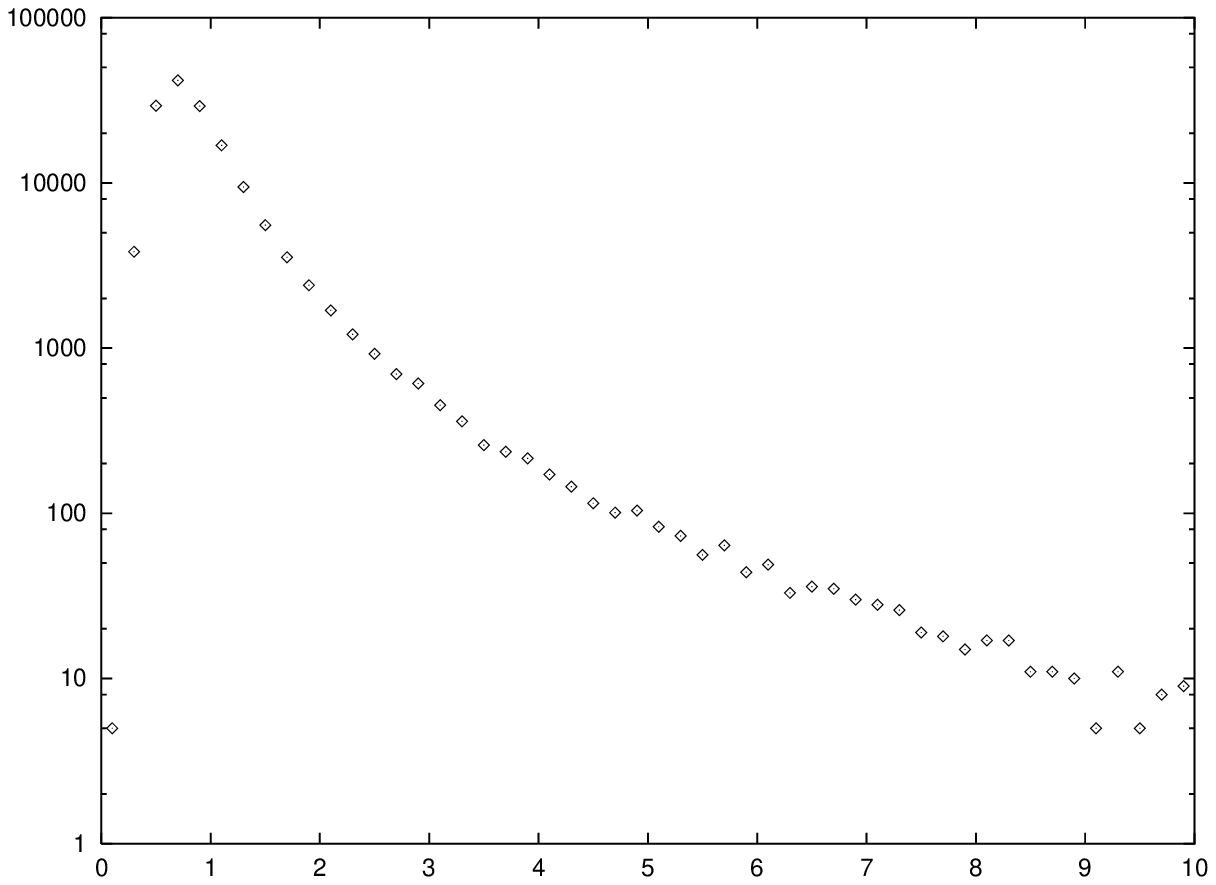}

\vspace{0.5cm}
\par
 {\bf Fig.3.} {\sl The rescaled weight distribution in a semi-log scale.}\\

\par
There are reasons for this  difference: e.g., we do not have a "coherence factor" $1-p$
decreesing the maximal possible value of weights even for big clusters of
particles with similar momenta values. Also we do not damp doubly the momentum
differences (by the definition of a cluster and by the prescription of weights).
We have only one free parameter instead of three as in Ref. [3]. This makes the
similarity of the results for the $Q^2$ and $m_W$ distributions quite
intriguing. 
\section {Summary and conclusions}
\par
We have applied the weight method to implement the Bose-Einstein interference
effect into the Monte Carlo generator for the $e^+e^-\to W^+W^-$ process
with a four jet final state. With the same value of the only free parameter
$\sigma $ as used for the $p \overline p$ we  get a reasonable qualitative
description of the Bose - Einstein ratio. The reconstructed value of the $W$
mass is practically the same with- and without weights. 
\par
One may argue that this last result is almost trivial: since the weights do not
change the momentum structure of events, there is no reason to expect mass
shifts for the reconstructed unstable particles. This is, however, a too
simplistic argument. Jet algorithms assign the decay products to the two $W$-s 
only statistically. One may easily imagine non-zero correlations between the
degree of misassignment and the weight values, which would introduce a signicant
mass shift. The absence of such shifts for two different weight methods supports
the suggestion that such correlations do not exist.  
\par
Summarizing, we confirm the claim of Ref. [3] that the seems to be no
significant $W$ mass shift in four jet events due to the Bose - Einstein
interference effect. Since our algorithm is based directly on the original
formula for weights [4] (apart from cutting the full series (3)) and contains 
no {\sl ad hoc} extra assumptions and free parameters, we regard it as a 
reliable, general method of implementing the Bose - Einstein effect. Therefore we
believe that our results are relevant for the coming precise $W$ mass
measurements.

\vspace{0.2cm}
{\large \bf Acknowledgements}
\vspace{0.2cm}
\par
A financial  support from KBN grants No 2 P03B 083 08 and No 2 P03B 196 09
is gratefully acknowledged. 
\vspace{1cm}

{\large \bf References}
\vspace{0.2cm}         

\par
\noindent 1. L. L\"onnblad and T. Sj\"ostrand, Phys. Lett. {\bf B351} (1995) 293.
\par
\noindent 2. J. Ellis and K. Geiger,  Phys. Rev.{\bf D 54} (1996) 1967.
\par
\noindent 3. S. Jadach and K. Zalewski,  Acta Phys. Pol. {\bf B 28} (1997) 1363.
\par
\noindent 4. A. Bia{\l}as and A. Krzywicki,  Phys. Lett. {\bf B 354} (1995) 134.
\par
\noindent 5. K. Fia{\l}kowski and R. Wit,  preprint TPJU-3/97, e-print
hep-ph/9703227, Z. Phys.  C, to be published.
\par
\noindent 6. D.H. Boal, C.-K. Gelbke and B.K. Jennings, Rev. Mod. Phys. {\bf 62}
(1990) 553, and references therein.
\par
\noindent 7. J.P. Sullivan at al., Phys. Rev. Lett.  {\bf 70} (1993) 3000.
\par
\noindent  8. B. Andersson and W. Hoffman, Phys. Lett. {\bf B169} (1986) 364,
 B. Andersson and M. Ringner, e-print hep-ph/9704383.
\par
\noindent 9. T. Sj\"ostrand and M. Bengtsson, Comp. Phys. Comm. {\bf 43} (1987) 367;
T. Sj\"ostrand, CERN preprint CERN-TH.7112/93 (1993), T. Sj\"ostrand and M. Bengtsson, 
Comp. Phys. Comm. {\bf 46} (1987) 43.
\par
\noindent 10. S. Pratt, Phys. Rev. Lett. {\bf 53} (1984) 1219.
\par
\noindent 11. J. Pi\u{s}\'{u}t, N. Pi\u{s}\'{u}tov\'{a}, B. Tom\'{a}\u{s}ik,
Phys.Lett. {\bf B368} (1996) 179; Acta Phys. Slov. {\bf 46} (1996) 517.
\par
\noindent 12. S. Haywood, Rutherford Lab. Report RAL-94-074 (1995).
\par
\noindent 13. J. Wosiek, Phys. Lett. {\bf B399} (1997) 130  
\par
\noindent 14. OPAL Collab., P.D. Acton et al., Phys. Lett. {\bf B267} (1991)
143, ALEPH Collab., D. Decamp et al.,  Z. Phys. {\bf C54} (1992) 75.  
\par
\noindent 15.  T. Sj\"ostrand, Comp. Phys. Com., {\bf 82} (1994) 74.
\end{document}